\documentclass[11pt]{article}
\pdfoutput=1
\usepackage{jcappub}
\usepackage{mathtools}
\usepackage{microtype}
\usepackage[utf8]{inputenc}
\usepackage[english]{babel}
\usepackage[dvipsnames]{xcolor}
\usepackage{amsmath,amssymb,amsbsy,amstext,amsthm}
\usepackage[colorlinks=true]{hyperref}
\usepackage{microtype}
\usepackage{graphicx}
\usepackage{amsfonts}
\usepackage{upgreek}
\usepackage{exscale,relsize}
\usepackage[makeroom]{cancel}
\usepackage{soul}
\usepackage{float}
\usepackage{empheq}
\usepackage[most]{tcolorbox}
\usepackage{subcaption}
\RequirePackage{color}
\usepackage{feynmp}
\usepackage{ifpdf}
\ifpdf
\fi

\usepackage{ragged2e}

\usepackage{colortbl}
\definecolor{green2}{cmyk}{0, 1, 0.5, 0}
\definecolor{lightgreen}{cmyk}{0.2, 0, 0.2, 0.2}
\definecolor{dred}{rgb}{0.9,0.2,0.5}
\definecolor{dred2}{cmyk}{0.1,0.7,0.1,0.3}
\definecolor{lightgray2}{cmyk}{0.4,0.4,0,0.8}
\definecolor{black}{cmyk}{1.0,1.0,1.0,1.0}

\definecolor{verde}{rgb}{0,0.5,0}

\AtBeginDocument{
 \hypersetup{
 urlcolor=NavyBlue,
 citecolor=NavyBlue,
 linkcolor=BrickRed,
 }
}

\allowdisplaybreaks[1]

\usepackage{colortbl}

\makeatletter
\newlength{\apb@width}
\newcommand{\autoparbox}[2][c]{\settowidth{\apb@width}{#2}\parbox[#1]{\apb@width}{#2}}

\makeatother

\numberwithin{equation}{section}

\def\beq{\begin{equation}}
\def\eeq{\end{equation}}

\def\bea{\begin{eqnarray}}
\def\eea{\end{eqnarray}}

\def\ni{\noindent}

\def\0{{\boldsymbol 0}}

\newtcbox{\mymath}[1][]{%
    nobeforeafter, math upper, tcbox raise base,
    enhanced, colframe=gray!30!gray,
    colback=gray!10, boxrule=0.5pt,
    #1}

\usepackage{setspace} 

\begin{document}

\title{Pure Natural Inflation Passes the ACT}

\author[\spadesuit,1]{Crist\'{o}bal Zenteno Gatica\note{corresponding author},}
\author[\spadesuit]{Matteo Fasiello,}
\author[\spadesuit]{and Alexandros Papageorgiou}

\affiliation[\spadesuit]{Instituto de F\'{i}sica T\'{e}orica UAM-CSIC, c/ Nicol\'{a}s Cabrera 13-15, 28049, Madrid, Spain}

\emailAdd{matteo.fasiello@csic.es}
\emailAdd{papageorgiou.hep@gmail.com}
\emailAdd{cristobal.zenteno@ift.csic.es}

\vspace{1.2cm}

\abstract{
\ni Pure natural inflation is a compelling effectively single-field model of inflation stemming from a top-down approach to the acceleration mechanism. In this short letter we show that such model is compatible with the latest CMB constraints obtained from the Atacama Cosmology Telescope combined with baryon acoustic oscillation data from the Dark Energy Spectroscopic Instrument. Under both the instantaneous reheating hypothesis and standard assumptions for reheating, we rule in a non-trivial fraction of the parameter space. We apply our analysis also to a phenomenological extension of the model and chart its viable parameter space.  
}

\maketitle


\section{Introduction}

\label{intro}

Whilst eagerly awaiting data from upcoming intermediate and small-scale probes accessing a completely new window on inflation, it is once again CMB data, when combined with DESI, that provide a new and possibly highly consequential  measurement \cite{AtacamaCosmologyTelescope:2025blo,Ferreira:2025lrd,McDonough:2025lzo}. A slightly larger value for the scalar spectral tilt $n_s$ has meant that a number of key compelling models including (the simplest realizations of) Starobinsky and Higgs inflation are now somewhat in tension with the data \cite{Zharov:2025zjg}. 
There are well-studied practical stratagems to resolve such tension, including non-minimal coupling to gravity \cite{Kallosh:2025rni}, modified gravity \cite{DOnofrio:2025bol},
and a stiff equation of state \cite{Haque:2025uga, Drees:2025ngb,Odintsov:2025bmp} during reheating. 

Under the assumption of  minimal coupling to gravity and standard post-inflationary evolution, in this brief letter we focus on a simple and natural \footnote{Natural in the sense that the inflaton is an axion like particle coupled to a pure Yang-Mills theory.} inflation model featuring
a significant portion of parameter space that is fully consistent with the latest ACT data. We do so first by assuming instantaneous reheating and later considering perturbative reheating.


\section{Pure Natural Inflation}
\label{PNI}
\begin{figure}[h]
  \centering
  \includegraphics[width=0.7\textwidth]{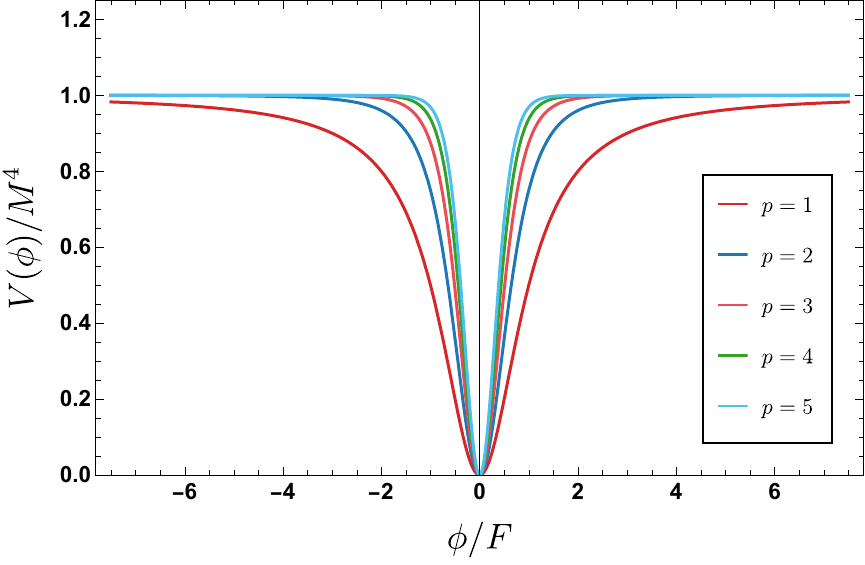}
  \caption{\justifying Plot of the pure natural potential Eq. (\ref{eqn:PureNaturalPot}) for several values of $p$.}
  \label{Fig:PNIpotential}
\end{figure}

The pure natural inflation (PNI) setup finds its  origin \cite{Nomura:2017ehb} as a model where an axion-like particle (ALP) is coupled to pure Yang-Mills theory via
\begin{equation}
    \label{eqn:PureYangMillsInt}
    \mathcal{L}_{} \supset \frac{1}{32 \pi^2}\frac{\phi}{f }\epsilon_{\mu\nu\rho\sigma}\,{\rm Tr}\, F_{\mu\nu}F_{\rho\sigma}\; ,
\end{equation}

\noindent where $f$ is the axion decay constant. In the large-$N$ description one introduces the 't~Hooft coupling
$ \tilde{\lambda}\equiv g^{2}N$. The term in Eq.~(\ref{eqn:PureYangMillsInt})  gives rise to a (multi-branch, reminiscent of the axion monodromy case, see e.g. \cite{McAllister:2008hb,McAllister:2014mpa,DAmico:2021vka,DAmico:2021fhz}) potential well-described in the large $N$ limit by (see Fig.~\ref{Fig:PNIpotential})  
\begin{equation}
    \label{eqn:PureNaturalPot}
    V(\phi) = M^4 \Bigg[1 - \frac{1}{\left( 1 + (\phi/F)^2 \right)^p}   \Bigg]\; ,
\end{equation}
valid for a single branch. Here $F\equiv 8\pi^2 N f/\tilde{\lambda}$ is the effective axion decay constant. We follow \cite{Nomura:2017ehb} in using  $\tilde{\lambda}\simeq 8\pi^{2}$ as a benchmark value for $\tilde{\lambda}$, implying $F\simeq Nf$. The value of $M$ is as usual determined by using the Friedman equation and the single-field slow-roll expression for the observed value of the power spectrum of scalar fluctuations. Given that $f$ must be sub-Planckian to keep the effective theory under control, a reasonable upper bound for $F$ is that of $({\rm a\,few}) \times M_{\rm p}$, while a $p$-dependent lower bound 
may be enforced upon requiring that inflation take place on a single branch \cite{Nomura:2017ehb}. Remarkably, even when limited to a single branch, the inflationary field excursion in PNI ($\Delta\phi \simeq F$) scales with an extra $N$ factor w.r.t.  well-known potentials such as e.g. the celebrated cosine of natural inflation \cite{Freese:1990rb}. This enables a large field excursion without resorting to a dangerously Planckian axion decay constant $f$. 

\begin{figure}[h]
  \centering
  \includegraphics[width=0.7\textwidth]{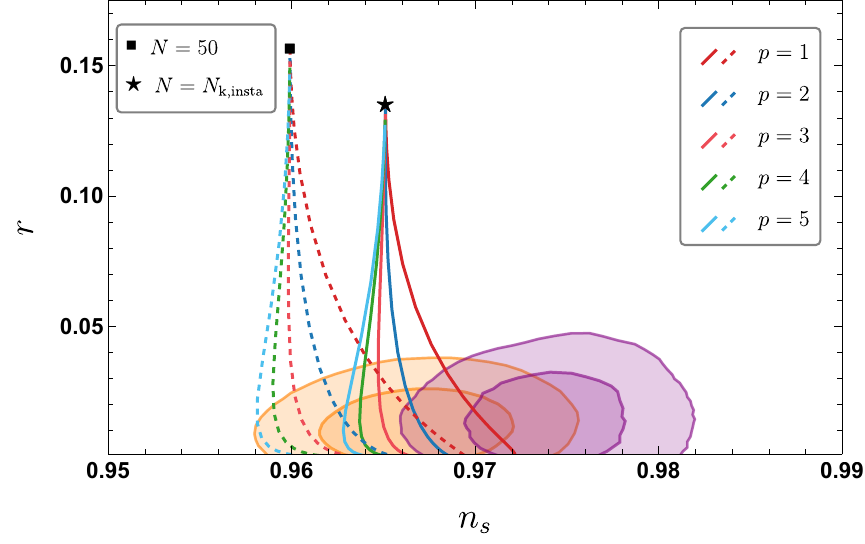}
  \caption{\justifying Trajectory traced in the $r$-$n_s$ plane as the parameter $F$ is varied in the pure natural inflation model. The light/dark orange contours correspond to Planck+BICEP constraints, while the purple regions stem from the Planck+ACT+CMB lensing+BK+DESI+BAO results ($n_s = 0.9743 \pm 0.0034$). Solid lines are computed from Eq.~(\ref{eqn:Nrh}) assuming instantaneous reheating, while the dashed lines assume an inflationary evolution of 50 e-folds. In both cases, for each $p$, the range of $F$ spans from $0.1 M_{\rm p}$ to $75M_{\rm p}$,  the latter value providing an excellent approximation for the predictions of the chaotic inflation scenario. These are identified by the black square or star, depending on the  duration of inflation.}
  \label{Fig:PlanckActContoursLinear}
\end{figure}

As shown in \cite{Nomura:2017ehb,Nomura_2018}, this model predictions do very well vis-\'{a}-vis  the Planck/Bicep constraints for $n_s$ and $r$ for a non-trivial fraction  of the parameter space ($F, p$). What we set out to do in this short note is checking how the same setup performs once the latest data from ACT \cite{AtacamaCosmologyTelescope:2025blo} is included. Upon varying the potential parameters $F$ and $p$, one can readily identify some clear patterns.\\
(i) With $F$ and $N$ fixed, increasing $p$  (specifically from $p=1$ to $p=10$) leads to a decrease in $n_s$. \\
(ii) For constant $p$ and $N$, an increasing $F$ generally raises $r$. At sufficiently large $F$ values the dynamics becomes independent of $p$ and converges towards the chaotic inflation case. The effect of a changing $F$ on $n_s$ is negligible for large $F$ and is otherwise highly dependent on $p$ (see Fig.~\ref{Fig:PlanckActContoursLinear}).\\

In Figure \ref{Fig:PlanckActContoursLinear} we show that, assuming instantaneous reheating, the values $p = {1,2}$ remain within the $2\sigma$ region of the ACT measurements\footnote{
As first stated in figure \ref{Fig:PlanckActContoursLinear}, whenever we mention ACT constraints on $r-n_s$ values or ACT data, we refer to the combined results from Planck+ACT+CMB lensing+BK+DESI+BAO ($n_s = 0.9743 \pm 0.0034$).
}. 
In this regime, a larger $n_s$ corresponds to smaller values of $F$ which are therefore favoured by the data. A smaller $F$ also corresponds to smaller values of the tensor to scalar ratio and brings the plateau region of the potential closer to the origin.\\
It is important to stress at this stage that one may not span arbitrarily small values of $F$. As explained in \cite{Nomura_2018}, too small an $F$ would e.g. significantly enhance the tunnelling probability between  adjacent branches \footnote{We are working under the hypothesis that full inflationary evolution takes place along one branch of the potential.}
Furthermore, the expression in Eq.~(\ref{eqn:PureNaturalPot}) is no longer necessarily a good description of the potential for $\phi\gtrsim N \pi f$. We limit the parameter space so as to never cross this threshold. Relatedly, if we set the moment when the CMB modes exit the horizon ($\phi_{\rm CMB}$) too far in the flat part of the potential, there is a chance that the evolution does not remain confined within the intended branch: it might shift to an adjacent one, invalidating our initial single-branch evolution assumption.

In \cite{Nomura_2018} the authors identify a criterion to avoid a change of branch. It is convenient to define  the dimensionless quantity $y \equiv \phi_*/F$, with $\phi_*$ the value of the field corresponding to the current horizon scale \footnote{We stress that, although close in value, $\phi_{\rm CMB}$ denotes the field value at the CMB pivot $k_{\rm CMB}=0.05\,{\rm Mpc}^{-1}$, while $\phi_\ast$ is the field value at the current horizon scale $k_0=a_0H_0$. As in \cite{Nomura_2018}, we apply the single-branch bound to $\phi_\ast$.}. The value of the control parameter $\gamma$ is determined from lattice analyses of such potentials \cite{Bonati:2016tvi} and can be expressed as a function of $p$: 

\begin{equation}
    \gamma(p) \simeq 0.47 \sqrt{p + 1}.
\end{equation}

\begin{figure}[h]
  \centering
  \includegraphics[width=0.7\textwidth]{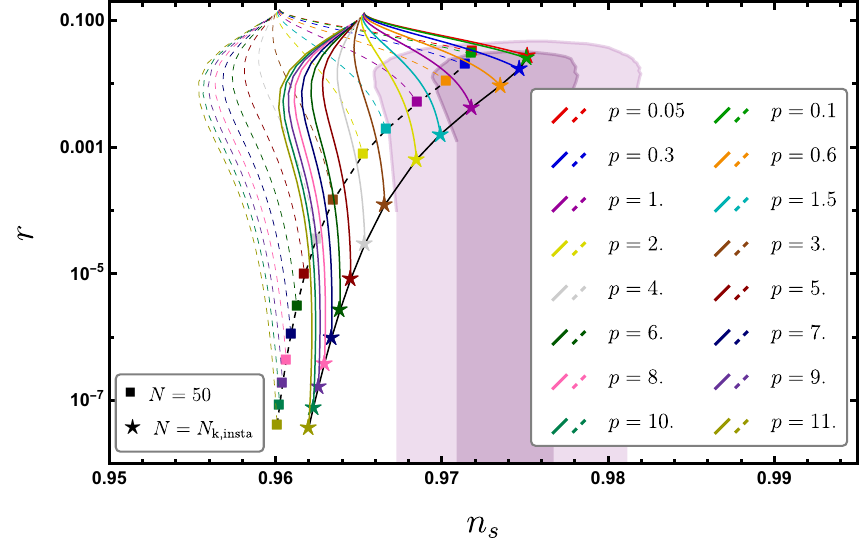}
  \caption{\justifying
  The $r-n_s$ plane is plotted here using a logarithmic scales for the $r$ axis in order to capture a wider range of values of the parameter space. The lowest value of $r$ corresponds to the lowest allowed value for $F$ and comes from the requirement that $\phi_*/F< 0.90/\gamma(p)$ where 90\% is a somewhat arbitrary threshold to make sure we are not too close to the actual bound \cite{Nomura_2018}. Dashed/Solid lines whose endpoint is a square/star are obtained as in Fig.~\ref{Fig:PlanckActContoursLinear}.}
  \label{fig:pExtNinsta}
\end{figure}

For a given $p$, a bound on $y$ straightforwardly translates into a constraint on $F$, identifying a value $F_{\rm min}$ and, correspondingly, setting a  lower bound on $r$. After accounting for  the $p$-dependent lower bound on $F$ in the allowed parameter space, one can see there  is in the $(r,n_s)$ plane a non-trivial region compatible with the ACT observations. As shown in Figure \ref{fig:pExtNinsta}, there is a clear preference for smaller values of $p$ when considering about  57 e-folds of inflationary evolution (the approximated range for instantaneous reheating obtained from  Eq. (\ref{eqn:Nrh})). If one would rather be conservative about re-heating and consider only 50 e-folds of inflation  (implementing  also the lower bound on $F$),  then it is the $p \leq 1$ region that lies within $2\sigma$ from the central values identified by ACT. We shall see later on that, under reasonable assumptions for perturbative reheating channels, the number of e-folds of inflation is never significantly smaller than 50.

The possibility of a small positive $p$, ($0<p<1$) is fully inline with the initial PNI proposal \cite{Nomura:2017ehb}.  
 Within this interval, extending the $F_{\min}$ constraint remains well motivated. Notably, for $p = 0.5$ and the corresponding value of $F_{\min}$, we find that the predictions for the spectral index and the tensor-to-scalar ratio under instantaneous reheating lie  close to the center of the ACT-preferred region, $(n_s = 0.974,\ r = 0.0011)$, see Figure \ref{fig:pSmallGG}. Along with the viability of this range in the case of instantaneous reheating, it is also noteworthy that only within this interval the $r-n_s$ results obtained assuming an inflationary evolution of 50 e-folds fall inside the $1\sigma$ region.

\section{Field Excursion}
\label{deltaPhi}
As a result of the presence of the number of colors parameter $N$, in PNI one finds an additional \footnote{Additional, for example, with respect to the  standard natural inflation model \cite{Freese:1990rb}.} handle to implement a trans-Planckian field displacement $\Delta \phi_{\rm eff}$. Crucially, the  microscopic axion excursion $\Delta\theta\sim \Delta \phi /(N f)$ remains order one \footnote{It's typically smaller than order one for the region of parameter space we probe here.} or smaller in PNI. This results in an effective theory that is typically under control, which requires $f \ll M_{\rm p}$, but can still deliver a large macroscopic field displacement.

In this work we do not straightforwardly implement any constraint stemming from the swampland distance conjecture (SDC, see e.g. \cite{Scalisi:2018eaz}). The SDC is a statement about geodesic motion in the genuine moduli space of quantum gravity predicting a tower of states with masses that can be dangerously light (close to $H$, in the case of inflation) for a large moduli-space displacement $\Delta s$. Whether there exists an exponentially light particle tower depends on the UV completion; an emergent $\theta$ from gauge dynamics might not be conducive to the existence of such tower. In particular, we are not aware of evidence that a displacement of an emergent $\theta$-angle in pure Yang-Mills is accompanied by an infinite tower of particles becoming parametrically light, although glueballs and domain walls may still play a role in the UV completion.

%
\section{Extended Model Space}
\label{ParameterSpace}

\begin{figure}[h]
  \centering
  \begin{subfigure}{\linewidth}
    \centering
    \includegraphics[width=0.7\textwidth]{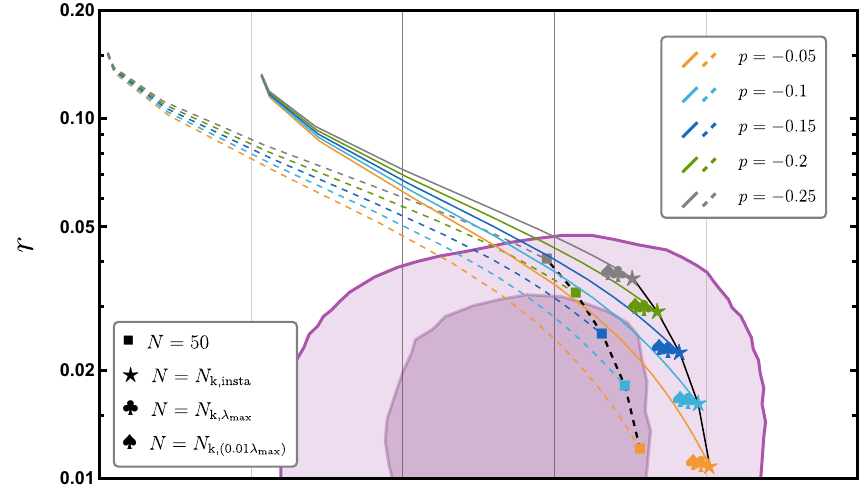}
  \end{subfigure}
  \begin{subfigure}{\linewidth}
    \centering
    \includegraphics[width=0.7\textwidth]{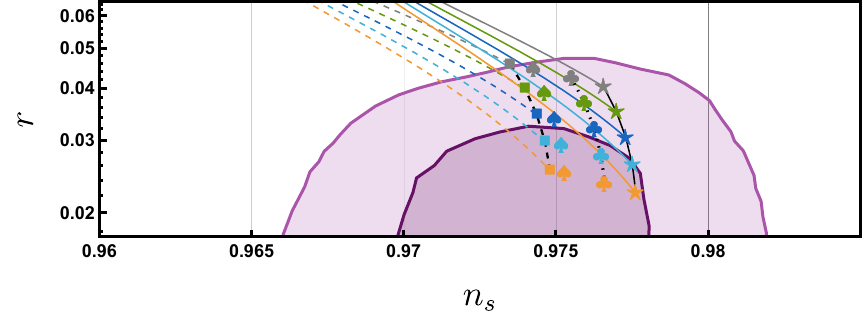}
  \end{subfigure}
  \caption{\justifying Results for small negative values of the parameter $p$. The upper panel tracks the evolution of observables as $F$ decreases from $75\; M_{\rm p}$ to $10^{-4}\; M_{\rm p}$, while the lower panel provides a detailed view for $F \geq 0.25 \;M_{\rm p}$ to highlight the range of parameter space consistent with $1\sigma$ observational constraints. Squares and stars follow the same conventions as in Fig.~\ref{fig:pExtNinsta}, denoting results for $N=50$ and instantaneous reheating, respectively. Intermediate points represent non-instantaneous reheating via the perturbative decay channel described in Sec. \ref{PostInfl}: clovers  denote the maximum coupling allowed by the backreaction constraint, while pikes correspond to $1\%$ of this maximum value.} 
  \label{fig:PnegPlots}
\end{figure}

The original PNI proposal posits a positive $p$ with the explored parameter space  typically corresponding to a $p$ with an integer value between 1 and 10. In what follows we shall adopt a more phenomenological perspective and investigate inflationary dynamics also for fractional and negative values of $p$. In order to make contact with the literature we should stress that a $-1<p<0$ in Eq.~(\ref{eqn:PureNaturalPot}) corresponds,  upon adding a sign flip, to potentials that have been used to model the phenomenology of monodromy-type potentials. These same potentials have been studied in the context of ALP dark matter in \cite{Chatrchyan:2023cmz} and are related to Eq.~(\ref{eqn:PureNaturalPot}) by the same $p$ sign flip. The family of potentials we will study in this section is then:

\begin{equation}
    \label{eqn:Sign}
    V_{p<0}(\phi) = {\rm Sgn}(p) \,M^4 \Bigg[1 - \frac{1}{\left( 1 + (\phi/F)^2 \right)^p}   \Bigg]\; ,
\end{equation}

An analysis  of the parameter space region with $p < 0$ shows that only values in the range $-0.25 < p < 0$ yield results consistent with the ACT contours. Larger negative $p$ values give too high a value for the tensor-to-scalar ratio. As shown in Figure \ref{fig:PnegPlots}, most of the selected parameter space lies within the $2\sigma$ region (in fact, the entire space does for an inflationary stage of at least $55$ e-folds). 

Another interesting result for this regime is that, unlike the positive $p$ case where $r$ can in principle be made arbitrarily small by lowering F, for negative $p$ the $r-n_s$ predictions approach an asymptote corresponding to the potential scaling as $V \sim \phi^{-2 p}$  at large field values.


\section{Post Inflationary Physics}
\label{PostInfl}

We have seen that the pure natural model of inflation remains compatible, over a wide range of the parameter space, with the constraints on the primordial power spectrum from the newly released ACT data. Naturally, post-inflationary evolution impacts CMB predictions so that one ought to clarify the reheating mechanism in place. Once inflation ends, the energy density stored in the inflaton must be transferred to relativistic degrees of freedom in order to initiate the radiation-dominated era. This process, together with the transition to thermal equilibrium, is referred to as reheating. In this work we will remain agnostic as to the microphysics that establish thermalization and instead focus on the primary goal of accounting for a radiation dominated universe. Depending on the structure and strength of the inflaton couplings, reheating may proceed perturbatively, or non-perturbatively via mechanisms such as parametric resonance and tachyonic preheating.

\begin{figure}[h]
  \centering
  \includegraphics[width=0.7\textwidth]{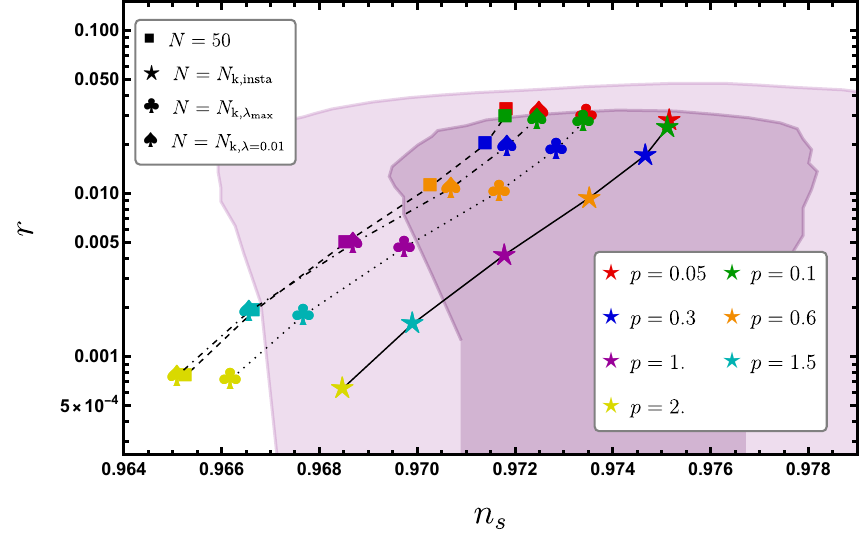}
  \caption{
  \justifying Zoomed-in version of Fig.~3 for positive values of the parameter $p$ in the range $p\in (0.05,2)$. The intermediate points correspond to the $(r,n_s)$ predictions obtained from the value of $N_k$ associated with a non-instantaneous reheating phase, as described in this section. Clover symbols, as in Fig.~4, denote the maximum coupling value allowed by the backreaction constraint, while pike symbols correspond to the coupling choice $\lambda = 0.01$.
  }
  \label{fig:pSmallGG}
\end{figure}

 Before a detailed discussion of the particular interaction chosen to model reheating in the pure natural inflation case, it is important to emphasize that the reheating phase is mainly characterized by three macroscopic parameters: its duration, the reheating temperature, and the effective equation-of-state parameter. If we use the approximation that the reheating period occurs with an approximately constant equation-of-state parameter, we can use matching equations \cite{Liddle:2003as} to derive a relation for the reheating duration:

\begin{equation}{\label{eqn:Nrh}}
N_{re} = \frac{4}{1 - 3 \omega}\left[
61.6 - \frac{1}{4}\ln\left(\frac{\rho_\phi(N_{\rm end})}{H(N_k)^4} \right)
-N_k
\right]\; ,
\end{equation}

where $N_k$ corresponds to the number of e-folds between the horizon exit of $k_{\rm CMB}$ ($0.05,\mathrm{Mpc}^{-1}$) and the end of inflation, and $\rho_\phi(N_{\rm end})$ is the energy density at the end of inflation. Using this equation, one can determine $N_k$ under the assumption of instantaneous reheating, which serves as a useful reference case. In Figure \ref{fig:pSmallGG} we show the $(r,n_s)$ predictions obtained under the assumption of instantaneous reheating for positive values of $p$, computed at the minimum allowed value of the parameter $F$ consistent with the constraint $\phi_*/F< 0.90/\gamma(p)$. In Figure \ref{fig:PnegPlots}, corresponding to the case of negative $p$, we also present the instantaneous reheating predictions for a selected set of $F$ values, reflecting the different phenomenological behavior in this part of the parameter space.

Given the shift symmetry in place for the axion, one of the simplest allowed \footnote{Note that the effect of a constant shift is simply a total derivative.} couplings is an axion–gauge field one, corresponding to the decay of the inflaton into Abelian gauge bosons,
\begin{equation}
    \mathcal{L}_{\rm int} = \frac{\lambda}{4 F}\phi\,  Y_{\mu\nu} \tilde{Y}^{\mu\nu},
    \label{CSc}
\end{equation}

\noindent with $Y$ the Abelian gauge field strength tensor and $\tilde{Y}$ its dual. The decay rate corresponding to this interaction term is given by  (see \cite{Weinberg_1996})
\begin{equation}
    \Gamma_{\phi \rightarrow \rm AA} = \frac{\lambda^2}{64 \pi}\frac{m_\phi^3}{F^2},
\end{equation}

\noindent with $m_\phi^2 = d^2 V/ d\phi^2$.

We ask that the coupling remain smaller than $\mathcal{O}(1)$ in order to preserve perturbative control. Furthermore, it is well known that axion–gauge field interactions of this type can induce significant backreaction effects in the gauge sector, potentially invalidating the perturbative treatment \cite{Anber:2009ua,Peloso:2013zw,vonEckardstein:2023gwk}. We therefore impose an additional upper bound by requiring the backreaction on the gauge fields  remain under control. In practice, we adopt the backreaction constraint derived for a $\mathrm{U}(1)$ axion–gauge field interaction  \cite{Anber:2009ua,Barnaby:2011vw}:
\begin{equation} \label{eqn:backreactionConstraint}
\frac{
\xi(\lambda)^{-3/2}e^{\pi\xi(\lambda)}
}{
79\frac{|\dot{\phi}|}{H^2}
}
\ll 1 ,
\end{equation}
\noindent where $\xi \equiv \lambda \dot{\phi}/(2 F H)$. The maximum value $\lambda_{\rm max}$ used in Figs. \ref{fig:PnegPlots} and \ref{fig:pSmallGG} for the computation of $r$ and $n_s$ is obtained by saturating the inequality in Eq. (\ref{eqn:backreactionConstraint}) and evaluating the dynamical quantities at the end of inflation. In the case of positive $p$, in addition to points corresponding to $\lambda_{\rm max}$ (typically of order unity), we also include the case $\lambda = 0.01$. For the negative-$p$ case, and for  selected values of $F$ and $p$ (see Fig.~\ref {fig:PnegPlots}), we also include an additional point corresponding to $\lambda = 0.01\,\lambda_{\rm max}$.

Qualitatively, a non-instantaneous standard (i.e. $0 \leq w_{\rm reh}\leq 1/3$) reheating phase shifts $N_k$ toward smaller values, leading to a smaller spectral index \footnote{In the numerical computation we assumed matter-like oscillations i.e. $\omega = 0$. We checked numerically, over the parameter space explored, that this assumption provides an accurate approximation.}. For positive $p$, this effect tends to move the $p=1$ predictions into the $2\sigma$ region of the ACT constraints, while the $p=2$ case becomes disfavored, see Figure \ref{fig:pSmallGG}. At the same time, values with $p<1$ remain compatible with observations within the $1\sigma$ contour. On the other hand, for negative $p$ this shift toward smaller values of $n_s$ is favored, as it moves the predicted $(r,n_s)$ points toward the $1\sigma$ region for several combinations of parameters, see Figure \ref{fig:PnegPlots}.

It is important to note that this class of potentials may, regardless of the inflaton decay channel,
support tachyonic instabilities during the post-inflationary evolution, potentially leading to non-perturbative reheating effects. This specific instability arises from the fact that the potential changes curvature along the inflationary trajectory. In particular, if inflation ends in a region of field space where the curvature is negative, the inflaton oscillations around the minimum necessarily involve repeated crossings of this negative-curvature region. Such crossings can tachyonically amplify field perturbations, opening the possibility of a tachyonic preheating–like reheating scenario.

We find that for negative $p$ values, and for the values of the parameter $F$ used  in Figure \ref{fig:PnegPlots}, the field value at the end of inflation lies well inside the negative-curvature region. As a consequence, the occurrence of non-perturbative effects in the immediate aftermath of inflation cannot be excluded in this region of the parameter space. A detailed analysis of this tachyonic instability and its impact on reheating dynamics lies beyond the scope of the present work.  Similar phenomena have been extensively studied in models with closely related phenomenology \cite{Tomberg:2021bll}. We should note that in the case of positive $p$, which comes with the additional constraint on the parameter $F$,  the field values at the end of the inflationary evolution lie within the positive-curvature region of the potential (or sufficiently close to the inflection point), thereby rendering such tachyonic effects subdominant or altogether absent.

Given the Chern-Simon type coupling in Eq.~(\ref{CSc}), there is of course also the possibility of tachyonic pre-heating (also known as gauge pre-heating) due to the instability of one polarization of the gauge fields. Modulo a mild dependence on the potential, direct comparison of the Chern-Simons couplings with the study in \cite{Adshead:2015pva} (see Fig.~13 therein) suggests that our parameter space, and in particular our values of the coupling ($\lambda/F\lesssim 7 M_{\rm p}^{-1}$),  do not support efficient gauge pre-heating. Similar reasoning applies to parametric resonance, which becomes inefficient as a reheating mechanism for small values of the coupling such as those we probed here. We stress that, for positive $p$ values, considering perturbative reheating is the most conservative choice in that it corresponds to the configuration that deviates the most from the region in the $(r,n_s)$ plane most favored by the ACT dataset. Any $p$ value  ruled in as a result of this work could only fare better under more efficient re-heating dynamics.


\section{Conclusions}
\label{conclusions}

In this work we revisited the Pure Natural Inflation model in light of the latest results from the Atacama Cosmology Telescope combined with baryon acoustic oscillation data from the Dark Energy Spectroscopic Instrument.
This scenario, compelling on its own, has recently \cite{Dimastrogiovanni:2025snj} been employed in the context of axion - gauge field models \cite{Anber:2009ua,Maleknejad:2011jw,Adshead:2012kp,Adshead:2013nka,Mukohyama:2014gba,Obata:2016tmo,Dimastrogiovanni:2016fuu,Garcia-Bellido:2016dkw,Thorne:2017jft,Agrawal:2017awz,Domcke:2018rvv,Holland:2020jdh,Caravano:2022epk,Figueroa:2023oxc,Garcia-Bellido:2023ser,Iarygina:2023mtj,Dimastrogiovanni:2023juq,Dimastrogiovanni:2023oid,Dimastrogiovanni:2024xvc,Figueroa:2024rkr,Jamieson:2025ngu,Ishiwata:2025wmo}. 

We explored the model predictions across the ($p,F$) parameter space
and presented the resulting trajectories in the $(n_s,r)$ plane.
We then included
post-inflationary evolution considering (i) instantaneous reheating, (ii) fixing the pivot scale to exit $N_k=50$
 e-folds before the end of inflation, and (iii) a simple perturbative reheating channel through an axion-gauge-field coupling.

Our main result is that PNI continues to provide an excellent fit to the latest
ACT constraints, with viable regions that remain compatible with  CMB
bounds while predicting potentially observable tensor modes. Notably, this
agreement is achieved within a particularly economical and natural framework: it does not rely on non-minimal couplings to gravity, nor does it require non-standard reheating assumptions. In this sense, PNI stands out as one well-motivated inflationary scenario that remains both theoretically controlled and observationally successful even when ACT data is combined with DESI.

We performed a similar analysis for a phenomenological extension of the model to negative $p$ values. We find that in this case small negative values with $|p|<1/4$ are compatible with the findings by ACT. 

\section{Acknowledgments}
It’s a pleasure to thank Ema Dimastrogiovanni and Alessandra Grieco for helpful discussions.~MF and AP acknowledge 
CSIC grant 20235AT005.~The work of  MF, AP, and CZ  is partially supported by the Spanish Research Agency (Agencia Estatal de Investigaci\'{o}n) through the Grant IFT Centro de Excelencia Severo Ochoa No CEX2020-001007-S, funded by MCIN/AEI/10.13039/501100011033.~Part of this work was carried out during the 2025 “The Dawn of Gravitational Wave Cosmology” workshop, supported by the Fundaci\'{o}n Ram\'{o}n Areces and hosted by the Centro de Ciencias de Benasque Pedro Pascual. We thank both the CCBPP and the Fundaci\'{o}n Areces for creating a stimulating and very productive research environment.

\addcontentsline{toc}{section}{References}
\bibliographystyle{utphys}
\bibliography{ref}

\end{document}